# Effects of exciton-plasmon strong coupling on third harmonic generation of WS$_2$ monolayer at periodic plasmonic interfaces

Maxim Sukharev[1,2] and Ruth Pachter[3]

[1]College of Integrative Sciences and Arts, Arizona State University, Mesa AZ 85201, USA

[2]Department of Physics, Arizona State University, Tempe AZ 85287, USA

[3]Materials and Manufacturing Directorate, Air Force Research Laboratory, Wright-Patterson Air Force Base, Dayton, OH 45433, USA

Abstract: We study theoretically optical response of WS$_2$ monolayer located near periodic metal nanostructured arrays in two and three dimensions. The emphasis of simulations is on the strong coupling regime of excitons supported by WS$_2$ and surface plasmon-polaritons supported by various periodic plasmonic interfaces. It is demonstrated that a monolayer of WS$_2$ placed in a close proximity of periodic arrays of either slits or holes results in a Rabi splitting of the corresponding surface plasmon-polariton resonance as revealed in calculated transmission and reflection spectra. The nonlinear regime, at which the monolayer of WS$_2$ exhibits the third harmonic generation is studied in details. It is shown that at the strong coupling regime the third harmonic signal is significantly affected by plasmon-polaritons and the symmetry of hybrid exciton-plasmon modes. It is also shown that the local electromagnetic field induced by plasmons is the major contributor to the enhancement of the third harmonic signal in three dimensions. The local electromagnetic fields resulted from the third harmonic generation are greatly localized and highly sensitive to the environment making it a great tool for nano-probes.

## Introduction

In our interest in exciton-plasmon phenomena, we note that recently transition metal dichalcogenide (TMD) monolayers became new building blocks in probing strong coupling between these nanomaterials and plasmonic structures. TMDs have a direct band-gap that is located energetically in the visible part of the spectrum, which is useful in utilizing surface plasmon-polariton states of metal nanostructures (such as periodic arrays of nano-holes or slits), thus enabling optical coupling to the two-dimensional materials. Strong exciton-plasmon coupling was suggested to potentially be useful for quantum processing information phenomena (see Refs. [1, 2] and references therein). For example, the large enhancement was applied to probe dark excitons in two-dimensional TMDs [3]. Moreover, these materials present a unique opportunity to study nonlinear optical phenomena at the nanoscale, owing to high third order nonlinearities exhibited by $WS_2$. For example, third harmonic generation (THG) has been observed for few-layer thin TMDs, e.g., $MoS_2$ and $WS_2$ [4, 5], and may be appropriate for probing phenomena of liquid exfoliated TMDs of varying thicknesses. We note that monolayer TMDs do not exhibit THG because they are non-centrosymmetric, but here we use the monolayer as an approximation to a thin TMD nanostructure.

The research in plasmonics is currently expanding to hybrid systems comprised of molecular excitons coupled to surface plasmon-polaritons.[6] Owing to the strong local field enhancement at the surface plasmon resonance the energy of the exciton-plasmon coupling may surpass all damping rates thus resulting in formation of hybrid exciton-plasmon states, which exhibit both molecular and plasmonic properties. The studies of such systems, in addition to fundamental understanding of light-matter interactions, also promise great advances in optical nano-devices and photonic computers.[7] The nonlinear plasmonics relating local plasmonic field enhancements as the main trigger for nonlinear effects that include high harmonic generation is currently experiencing significant growth.[8, 9] Our goal in this paper is to combine the high third order nonlinearity of a $WS_2$ monolayer with the local plasmon field enhancement in the strong coupling regime.

In this manuscript we investigate for the first time how exciton-plasmon strong coupling influences the nonlinear optical response of hybrid nanomaterials, namely how the third harmonic generation by a $WS_2$ monolayer is affected by resonant plasmonic systems in two- and three dimensions.

## Model

The optical response is simulated using Maxwell's equations in the form

$$\frac{\partial \vec{B}}{\partial t} = -\nabla \times \vec{E},$$
$$\frac{\partial \vec{D}}{\partial t} = \frac{1}{\mu_0} \nabla \times \vec{B}, \quad (1.1)$$

where the electric displacement field is $\vec{D} = \varepsilon_0 \vec{E} + \vec{P}$.

The response of the metal is taken into account using the Drude-Lorentz model with the dielectric function reading as

$$\varepsilon(\omega) = 1 - \frac{\Omega_p^2}{\omega^2 - i\omega\Gamma} + \sum_n \frac{f_n \omega_n^2}{\omega_n^2 - \omega^2 + i\omega\gamma_n} \quad (1.2)$$

and the following set of fitting parameters: $\Omega_p$ (the plasma frequency), $\Gamma$ (the Drude damping), $f_n$ ($n^{th}$ Lorentz oscillator strength), $\omega_n$ (frequency of the $n^{th}$ Lo-rentz oscillator), and $\gamma_n$ (the Lorentz damping parameter for the n$^{th}$ oscillator), which are taken from Ref. [10] for silver and gold.

The linear optical response of monolayer WS$_2$ is calculated using the Lorentz dielectric function with five poles using parameters measured in Ref. [11][1]. The third order nonlinearity of WS$_2$ is simulated as an instantaneous Kerr effect with the isotropic in-plane third-order susceptibility $\chi^{(3)} = 2.611 \times 10^{-17}$ m$^2$/V$^2$ [5].

The Maxwell equations (1.1) are discretized in space and time along with the corresponding equations on current density governed by the material dispersion relation (1.2) in spatial regions occupied by metal. Similarly, the current density associated with the dispersion of WS$_2$ is also included in the simulations in both linear and nonlinear regimes. The resulting system of equations is propagated in time and space using the finite-difference time-domain (FDTD) approach [12]. We consider periodic systems in two and three dimensions under normal incidence conditions. The open boundaries are terminated by convolutional perfectly matched layers [13]. When simulating the nonlinear response we include the WS$_2$ dispersion and follow the fully explicit numerical algorithm proposed in Ref. [14], further developed and validated in Ref. [15]. We found that this algorithm leads to the same results as obtained by the commonly used GVADE method by Greene and Taflove [16]. The former however requires significantly less numerical effort (especially in three dimensions). The details of the numerical approach used in this manuscript can be found in Ref. [15]. The numerical convergence in both linear and nonlinear regimes is achieved for the spatial resolution of 3.0 nm and the time step of 0.005 fs. The codes are parallelized to speed up calculations. The simulations were performed at the AFRL HPC center and at the Arizona State University local multiprocessor cluster.

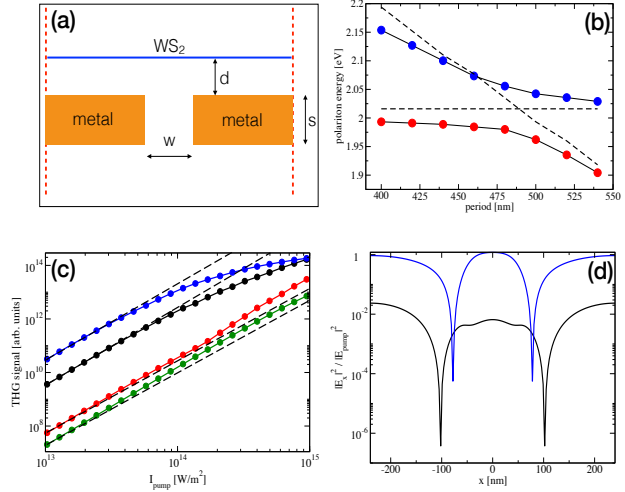

Fig. 1. Summary of a two-dimensional simulations. Panel (a) shows a schematic setup of two-dimensional array of slits with a monolayer of WS$_2$. The periodic array of slits has a period $p$; the thickness of the silver film, $s$, is 150 nm; the slit's width, $w$, is 100 nm. The WS$_2$ monolayer is placed at a distance $d$ above the metal. Panel (b) shows the energies of the upper and lower polaritons (blue and red circles, respectively) as a function of the period, $p$. The dashed lines picture $\omega_A$, corresponding to the frequency of the direct band gap transition in WS$_2$ (2.02 eV, horizontal dashed line), and the SPP mode. Panel (c) shows the THG signal as a function of the pump intensity calculated for the period of 480 nm. The black and blue lines show data obtained for the pump frequencies corresponding to the lower and upper polariton energies (1.98 eV and 2.06 eV), respectively. The red line shows data for the pump frequency $\omega_A$ and the green line data for the free-standing WS$_2$ (no metal) pumped at $\omega_A$. The parallel dashed lines show (I$_{pump}$)$^3$ dependence. Panel (d) shows the square of the horizontal component of the local electric field at the position of the WS$_2$ monolayer for the lower (black line) and upper (blue line) polaritons as a function of the horizontal coordinate $x$. Both distributions are calculated at pump frequencies that correspond to either lower or upper polaritons.

---

[1] The supplemental material of Ref. [2] provides numerical parameters for all Lorentz poles. The table S1 contains a typo – oscillator strengths for all poles must be divided by the factor of 4 in order to obtain the refractive index seen in Fig. S2b.

The linear response for all geometries considered is evaluated using a short-pulse method when both transmission and reflection spectra are obtained within a single FDTD run. When simulating THG, much attention was paid to achieving a proper steady-state regime, i.e. when the THG signal scales as a cube of the pump intensity. It was found that a total propagation time of 2 ps was required to achieve the steady-state regime for both two- and three-dimensional examples considered here. The THG signal is calculated as follows: first, the steady-state regime is achieved; next, the scattered fields corresponding to the third harmonic are obtained in the far-field zone along a contour spatially enclosing the system; finally, the outgoing energy flux (the Poynting vector) is integrated along a contour in the far-field zone.

## Results

First we consider the geometry shown in Fig. 1a. A periodic array of slits in a silver film with parameters provided in the figure caption is excited by a horizontally polarized plane wave at normal incidence. A $WS_2$ monolayer is located on the input side at a distance $d = 20$ nm above the array. It has been shown experimentally [11] that monolayer $WS_2$ exhibits an excitonic transition with a sharp photoluminescence line at $\omega_A = 2.02$ eV. The periodicity of a slit array can be accordingly adjusted to tune the energy of its surface plasmon-polariton (SPP) resonance, namely to the excitonic transition. For material parameters shown in Fig. 1 the slit array with a period of 480 nm supports the SPP mode at 2.02 eV.

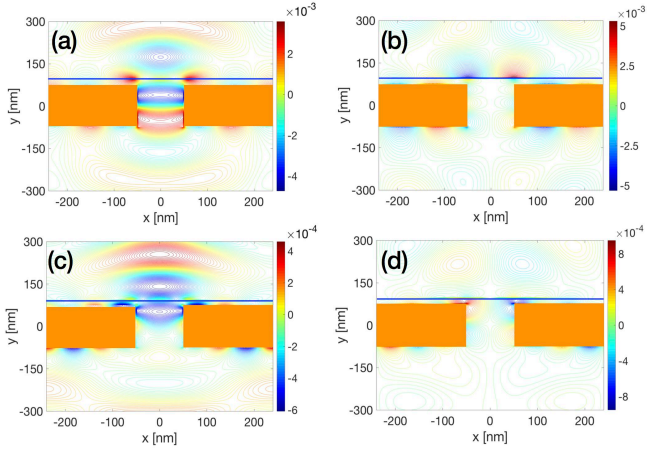

Fig. 2. Spatial distributions of the THG electric field calculated for pump frequencies corresponding to the lower polariton (panels (a) and (b)) and the upper polariton (panels (c) and (d)). The horizontal component of the electric field is shown in panels (a) and (c) and the vertical components are shown in panels (b) and (d). The field amplitude is normalized with respect to the pump amplitude. The pump intensity is $1.4 \times 10^{13}$ W/m². The orange rectangles represent metal and the horizontal blue line the spatial position of two-dimensional $WS_2$.

To demonstrate that the observed exciton-plasmon coupling is due to hybridization of the plasmon mode with the corresponding exciton mode of the monolayer, we performed a period sweep, while measuring the energy positions of the upper and lower polaritons in the transmission spectra. The results are shown along with the bare SPP mode, and the frequency of the direct band gap excitonic transition in Fig. 1b. A clear avoided crossing is observed with the lower polaritonic branch, having characteristics of the plasmon mode, while the upper polariton exhibits excitonic behavior both at high periods or for a small in-plane propagation constant. The calculated Rabi splitting for the period of 480 nm is 77 meV.

We now turn to the nonlinear response of the two-dimensional hybrid system. We examine the THG signal calculated as a function of the pump intensity at different pump frequencies for the period of 480 nm corresponding to the strongest hybridization (Fig. 1b). In order to better understand the influence of surface plasmons on the THG signal, it is also calculated for a free-standing $WS_2$ monolayer at a pump frequency of 2.02 eV ($\omega_A$), as shown as green line in Fig. 1c. The three other THG signals are: the black line corresponds to a pump frequency of 1.98 eV (the frequency of the lower polariton); red line shows THG for a pump frequency of 2.02 eV ($\omega_A$); and the blue line shows the THG calculated at a pump frequency of 2.06 eV (frequency of the upper polariton). Other parameters are provided in the figure caption.

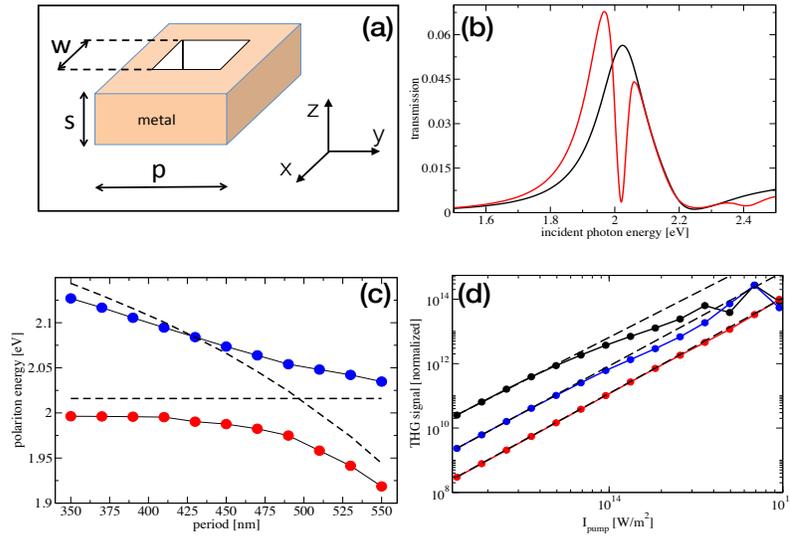

Fig. 3. Summary of three-dimensional simulations. Panel (a) shows schematic setup depicting a periodic array of square holes in 150 nm thin gold film. The periodicity is varied between 350 nm and 550 nm. The square holes are 140 nm on a side. The monolayer of $WS_2$ is placed on the input side 20 nm above the array. The system is excited by an x-polarized plane wave at normal incidence propagating in the negative z-direction. Panel (b) shows linear transmission as a function of the incident photon energy for the bare hole array for a period of 490 nm (black line) and for the hybrid system with a $WS_2$ monolayer (red line). Panel (c) shows the energies of the upper and lower polaritons (blue and red circles, respectively) as functions of the period, $p$. The dashed lines depict the frequency, at which the stand-alone $WS_2$ has a maximum absorption, $\omega_A$, (2.02 eV, horizontal line), and the SPP mode as a function of the period. Panel (d) shows the THG signal as a function of the pump intensity calculated for the period of 490 nm. The black and blue circles show data obtained for pump frequencies corresponding to the lower and upper polariton energies (1.97 eV and 2.05 eV). The red circles show data for the pump frequency corresponding to $\omega_A$. The parallel dashed lines show $(I_{pump})^3$ dependence.

All curves in Fig. 1c scale as $I_{pump}^n$, where $n \approx 2.9$ below $10^{14}$ W/m², indicating that the third harmonic regime is clearly achieved. At high pump intensities however both red and green curves deviate from the cubic dependence exhibiting higher $n$. With the pump intensities above $10^{15}$ W/m² these curves exhibit saturation. We attribute such an odd behavior to the numerical convergence. We found it notoriously difficult to numerically converge simulations (to achieve the steady-state regime) in both two and three dimensions at the pump frequency corresponding to $\omega_A$. In order to avoid numerical artifacts we thus discuss in the following pump intensities at which the cubic dependence of THG is clearly obtained.

It is evident that the presence of the metal significantly enhances the THG signal for all pump frequencies. When the system is pumped at either the lower or upper polaritonic frequency we observe a noticeable enhancement in THG, increasing several orders of magnitude as compared to pumping at a frequency of 2.02 eV. Furthermore, the THG signal begins to saturate at high pump intensities for all pump frequencies. However, the saturation is more dramatic for a pump frequency corresponding to the upper polariton. Fig. 1d compares horizontal spatial variations of the squared in-plane electric field for two hybrid states, noting that the local field enhancement for the upper polariton is significantly higher than that for the lower polariton, leading to higher THG signals at lower pump intensities.

It has been recently suggested that high harmonic local electric fields can be utilized to alter molecules placed at metal interfaces.[9] This is in part due to high spatial localization of the fields and their relative high amplitudes. It is thus informative to examine how such fields vary with the pump frequency and spatial coordinates. Fig. 2 shows results for the pump frequencies, corresponding to the lower and upper polaritonic states. The pump intensities for each pump frequency is $1.4 \times 10^{13}$ W/m$^2$ corresponding to a nearly perfect cubic dependence of THG on the pump intensity (see Fig. 1c). We note that even though the electric field at $\omega = 3\omega_{pump}$ has both horizontal and vertical components, only the horizontal component is being generated by the monolayer of WS$_2$ due to the dimensionality of the problem. The vertical component is in-duced solely by the slit array. Clearly the THG signal is mainly reflected by the metal film, with just a small fraction making past the slit array propagating in the negative y-direction. The fields shown in Fig. 2 exhibit complex spatial variations with several hot spots located near and inside slits. It is also important to emphasize that the variations of the phase of each component is highly spatially dependent. For exam-ple, the horizontal component for the third harmonic of the lower polariton (Fig. 2a) exhibits a π phase change within a distance of less than 80 nm, which is smaller than the wavelength corresponding to the third harmonic (208 nm in case of the lower polariton).

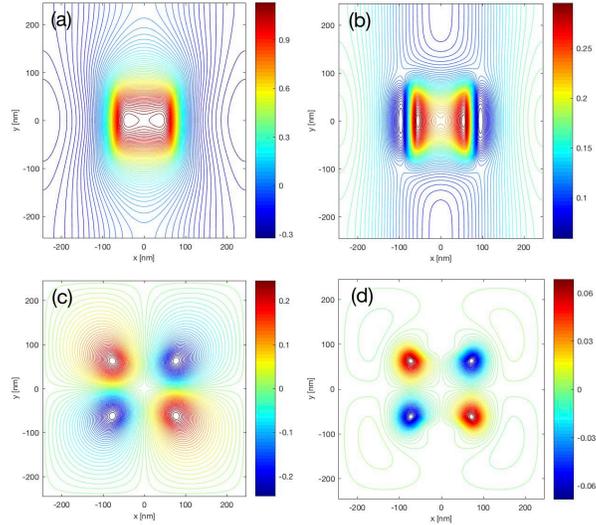

Fig. 4. Spatial distributions of electric field components for the hybrid system of a coupled hole array with a monolayer of WS$_2$. The fields are evaluated at the plane of WS$_2$ (20 nm above the metal). Panels (a) and (c) show the x-component of the electric field calculated for the frequency of the lower (1.97 eV) and upper (2.05 eV) polaritons, respectively. Corresponding induced in-plane y-components are shown in panels (b) and (d). The period of the array is 490 nm. The field amplitude is normalized to that of the incident field (i.e. units of enhancement).

The second study case we considered here is a three-dimensional setup schematically depicted in Fig. 3a. Here we consider a periodic array of square holes in a thin gold film with geometrical parameters close to Ref. [11]. The system is excited by a linearly polarized plane wave propagating in the negative z-direction. Typical linear transmission spectra are shown in Fig. 3b with $WS_2$ placed 20 nm above the metal on the input side (red line). Similar to the example of slit arrays we observe a Rabi splitting of about 80 meV for a period of 490 nm. We also note that the transmission for the hybrid system noticeably differs from that calculated for the hole array at higher energies around 2. 4 eV. This is due to the fact that the monolayer of $WS_2$ exhibits another dipolar resonance (B peak, due to spin-orbit coupling) at these energies, Ref. [11]), although less pronounced and with significantly higher losses compared to the direct band gap excitonic transition. Fig. 3c shows the avoided crossing behavior of the hybrid states of the hole array coupled to $WS_2$ monolayer as a function of the periodicity. The main results of the simulations presenting THG signals, calculated at different pump frequencies as functions of the pump intensity, are shown in Fig. 3d. THG signal scales as $I_{pump}^n$ with $n \approx 2.95$ below $7 \times 10^{13}$ W/m² for all three pump frequencies. It begins to saturate near $10^{14}$ W/m².

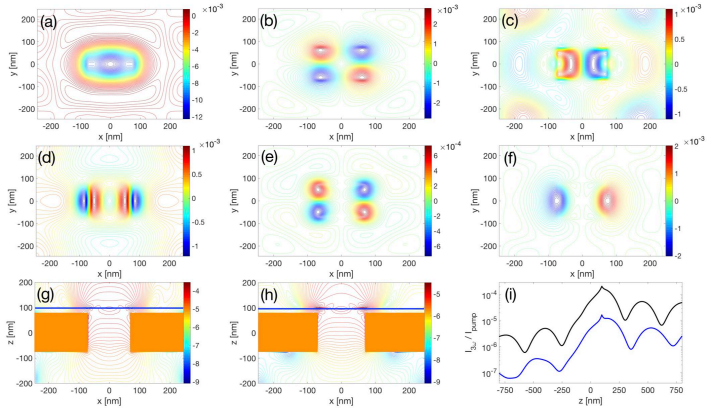

Fig. 5. Spatial distributions of the electric field associated with THG for the periodic array of holes with $WS_2$ at a distance of 20 nm above the metal. The upper raw (panels (a), (b), and (c)) corresponds to $E_x$, $E_y$, and $E_z$ components for the pump frequency of 1.97 eV (lower polariton). The second raw (panels (d), (e), and (f)) shows field components for the pump frequency of 2.05 eV (upper polariton). Both sets of fields is calculated at a distance of 10 nm above the metal. Panel (g) shows the intensity of the third harmonic as a function of x and z for 1.97 eV and panel (h) is for 2.05 eV – both are in log scale. Panel (i) compares the intensities of the third harmonic at two pump frequencies as functions of z-coordinate at x=y=0. All fields are calculated at the pump intensity of $1.4 \times 10^{13}$ W/m² and normalized with respect to the pump amplitude (in units of enhancement).

It is interesting to qualitatively compare the THG results in two and three dimensions, as in 3D the excitonic monolayer becomes a 2D object and is thus subjected to both $E_x$ and $E_y$ that contribute to THG. Periodic arrays of holes of various configurations present a unique opportunity to study the influence of the induced in-plane electric field caused by surface plasmon-polaritons on the nonlinear optical response of hybrid systems [17]. Comparing the results in Fig. 3d and Fig. 1c, we note that the three-dimensional geometry favors the lower polariton with THG signals that are prominently higher compared to the upper polariton. This behavior is opposite to that of the slit arrays geometry, where the THG signal is considerably higher for the upper polariton. We explain this phenomenon by the fact that the x-polarized incident pump is supplemented by the in-plane y-component of the electric field induced by the surface plasmon. Quantitatively when comparing THG signals in two- and three-dimensions we see that the THG signal calculated for the upper polariton for the hole array (Fig. 3d, black line) is noticeably higher than its two-dimensional analogue (Fig. 1c, black line), while the lower polaritons (corresponding blue lines) in both cases are quite similar in magnitude.

To support this idea, we calculate spatial distributions of in-plane electric field components corresponding to the lower and upper polaritons (Fig. 4). The fields are recorded at the plane where the monolayer of $WS_2$ is located. We note that the y-component of the electric field is of the same order as $E_x$, and it also exhibits a quadrupole type pattern. Moreover, the amplitude of the in-plane field for the lower polariton (Fig. 4a, c) is higher by the factor of 4 compared to that for the upper polariton (Fig. 4b, d), which explains why the periodic array of square holes generates more efficiently the third harmonic.

Next, we examine the spatial distribution of the fields associated with the third harmonic at different pump frequencies, namely of the lower and upper polaritons. The results of the simulations are presented in Fig. 5. Pumping the lower polariton (panels (a) through (c)) results in a higher local in-plane electric field, as compared to the upper polariton (panels (d) through (f)), as we have anticipated. Furthermore, the quadrupole pattern seen in Fig. 4 is clearly demonstrated in panels (b) and (e). It is also important to note that there is a significant lateral component, $E_z$, of the third harmonic for both pump frequencies, which is obviously generated by the metal. It is highly localized near the holes with considerable spatial variations, especially for the lower polariton (panel (c)). The spatial distributions of the intensities in the lateral direction (panels (g) through (i)) demonstrate that the third harmonic is highly spatially localized in the region near the holes with an oscillating pattern seen on the input and output sides of the metal. The intensities at both pump frequencies are localized within a region of about 100 nm near the monolayer of $WS_2$.

## Conclusion

We investigated for the first time how the exciton-plasmon strong coupling influences the nonlinear optical responses of hybrid nanomaterials. The emphasis of simulations was on the strong coupling regime of excitons supported by $WS_2$ and surface plasmon-polaritons supported by periodic plasmonic interfaces. Our major goal was to understand how the process of the third harmonic generation by a $WS_2$ monolayer might be affected by resonant plasmonic systems in two- and three dimensions. We demonstrated that a monolayer of $WS_2$ placed in a close proximity of periodic arrays of either slits or holes results in a Rabi splitting of the corresponding surface plasmon-polariton resonance as revealed in calculated transmission and reflection spectra. The nonlinear regime, at which the monolayer of $WS_2$ exhibits the third harmonic generation was studied in details. We showed that at the strong coupling regime the third harmonic signal is significantly affected by plasmon-polaritons and the symmetry of hybrid exciton-plasmon modes. It was also shown that the local electromagnetic fields induced by plasmons are the major contributor to the enhancement of the third harmonic signal in three dimensions. Furthermore the electromagnetic fields associated with the third harmonic signal are highly localized in space exhibiting local hot spots making it a great tool for future nano-probes.

## Acknowledgements

The financial and computing support of this work is provided by the Air Force Office of Scientific Research under Grant No. FA95501510189. MS would also like to thank Binational Science Foundation for generous financial support through Grant No. 2014113.

## References

[1] D. Zheng, S. Zhang, Q. Deng, M. Kang, P. Nordlander, H. Xu, Manipulating Coherent Plasmon–Exciton Interaction in a Single Silver Nanorod on Monolayer WSe2, Nano Lett, 17 (2017) 3809-3814.
[2] J. Wen, H. Wang, W. Wang, Z. Deng, C. Zhuang, Y. Zhang, F. Liu, J. She, J. Chen, H. Chen, S. Deng, N. Xu, Room-Temperature Strong Light–Matter Interaction with Active Control in Single Plasmonic Nanorod Coupled with Two-Dimensional Atomic Crystals, Nano Lett, 17 (2017) 4689-4697.
[3] Y. Zhou, G. Scuri, D.S. Wild, A.A. High, A. Dibos, L.A. Jauregui, C. Shu, K. De Greve, K. Pistunova, A.Y. Joe, T. Taniguchi, K. Watanabe, P. Kim, M.D. Lukin, H. Park, Probing dark excitons in atomically thin semiconductors via near-field coupling to surface plasmon polaritons, Nat Nano, 12 (2017) 856-860.
[4] R. Wang, H.-C. Chien, J. Kumar, N. Kumar, H.-Y. Chiu, H. Zhao, Third-Harmonic Generation in Ultrathin Films of MoS2, Acs Appl Mater Inter, 6 (2014) 314-318.
[5] T.-T. Carlos, P.-L. Néstor, E. Ana Laura, R.G. Humberto, A.C. David, B. Ayse, L.-U. Florentino, T. Humberto, T. Mauricio, Third order nonlinear optical response exhibited by mono- and few-layers of WS 2, 2D Materials, 3 (2016) 021005.
[6] M. Sukharev, A. Nitzan, Optics of plasmon-exciton nanomaterials, J Phys Condens Matter, DOI 10.1088/1361-648X/aa85ef(2017).
[7] P. Törmä, W.L. Barnes, Strong coupling between surface plasmon polaritons and emitters: a review, Reports on Progress in Physics, 78 (2015) 013901.
[8] M. Kauranen, A.V. Zayats, Nonlinear plasmonics, Nat Photon, 6 (2012) 737-748.
[9] J. Butet, P.-F. Brevet, O.J.F. Martin, Optical Second Harmonic Generation in Plasmonic Nanostructures: From Fundamental Principles to Advanced Applications, Acs Nano, 9 (2015) 10545-10562.
[10] A.D. Rakic, A.B. Djurisic, J.M. Elazar, M.L. Majewski, Optical properties of metallic films for vertical-cavity optoelectronic devices, Appl Optics, 37 (1998) 5271-5283.

[11] S. Wang, S. Li, T. Chervy, A. Shalabney, S. Azzini, E. Orgiu, J.A. Hutchison, C. Genet, P. Samorì, T.W. Ebbesen, Coherent Coupling of WS2 Monolayers with Metallic Photonic Nanostructures at Room Temperature, Nano Lett, 16 (2016) 4368-4374.
[12] A. Taflove, S.C. Hagness, Computational electrodynamics : the finite-difference time-domain method, 3rd ed., Artech House, Boston, 2005.
[13] S.D. Gedney, G. Liu, J.A. Roden, A.M. Zhu, Perfectly matched layer media with CFS for an unconditionally stable ADI-FDTD method, Ieee T Antenn Propag, 49 (2001) 1554-1559.
[14] D.F. Gordon, M.H. Helle, J.R. Peñano, Fully explicit nonlinear optics model in a particle-in-cell framework, J Comput Phys, 250 (2013) 388-402.
[15] C. Varin, R. Emms, G. Bart, T. Fennel, T. Brabec, Explicit formulation of second and third order optical nonlinearity in the FDTD framework, arXiv preprint arXiv:1603.09410, DOI (2016).
[16] J.H. Greene, A. Taflove, General vector auxiliary differential equation finite-difference time-domain method for nonlinear optics, Opt Express, 14 (2006) 8305-8310.
[17] F.J. García de Abajo, Colloquium : Light scattering by particle and hole arrays, Rev Mod Phys, 79 (2007) 1267-1290.